# From Continuous to First-Order-Like: Amorphous-to-Amorphous Transition in Phase-Change Materials

Tomoki Fujita[1]*, Yoshio Kono[2], Yuhan Chen[3], Jens Moesgaard[1], Seiya Takahashi[4], Arune Makareviciute[1], Sho Kakizawa[5], Davide Campi[6], Marco Bernasconi[6], Koji Ohara[7], Ichiro Inoue[8], Yujiro Hayashi[8], Makina Yabashi[8], Eiji Nishibori[4], Riccardo Mazzarello[3]*, Shuai Wei[1,9]*

[1]Department of Chemistry, Aarhus University, 8000 Aarhus C, Denmark.
[2]Department of Physics and Astronomy, Kwansei Gakuin University, Sanda 669-1330, Japan.
[3]Department of Physics, Sapienza University of Rome, Rome 00185, Italy.
[4]Department of Physics, Faculty of Pure and Applied Sciences and Tsukuba Research Center for Energy Materials Science (TREMS), University of Tsukuba, Ibaraki 305-8571, Japan.
[5]Japan Synchrotron Radiation Research Institute, 1-1-1 Kouto, Sayo-cho, Sayo-gun, Hyogo 679-5198, Japan.
[6]Department of Materials Science, University of Milano-Bicocca, I-20125 Milano, Italy.
[7]Faculty of Materials for Energy, Shimane University, Matsue, Shimane 690-8504, Japan.
[8]RIKEN SPring-8 Center, 1-1-1 Kouto, Sayo-cho, Sayo-gun, Hyogo 679-5148, Japan.
[9]iMAT Centre for Integrated Materials Research, Aarhus University, Denmark.

Polymorphism is ubiquitous in crystalline solids. Amorphous solids, such as glassy water and silicon, may undergo amorphous-to-amorphous transitions (AATs). The nature of AATs remains ambiguous, due to diverse system-dependent behaviors and experimental challenges to characterize disordered structures. Here, we identify two ordered motifs in amorphous phase-change materials and monitor their interplay upon pressure-induced AATs. Tuning temperature, we find a crossover from continuous to first-order-like AATs. The crossover emerges at a special pressure-temperature combination, where the AAT encounters a maximum in crystallization rate. Analyzing the two ordered motifs in a two-state model, we draw a phenomenological parallel to the phase transition behavior of supercooled water near its second critical point. This analogy raises an intriguing question regarding the existence of a critical-like point within amorphous solids.

## Introduction

Polymorphism is a well-known phenomenon in crystalline solids referring to the existence of two or more crystal structures with the same chemical composition. Likewise, the terms "liquid polymorphism" and "polyamorphism" refer to the existence of two or more states with distinct (disordered) structure and properties in non-crystalline materials such as liquids and glasses(*1, 2*). The corresponding structural transitions are called liquid-liquid transitions (LLTs) and amorphous-amorphous transitions (AATs)(*3*), respectively. The LLTs are typically associated with anomalous pressure (*P*)- temperature (*T*) responses in the thermodynamic properties (e.g. maxima in heat capacities and density anomalies(*4*)) or in the dynamic properties (e.g. fragile-to-strong transition(*5*)), which are not fully understood. Much effort has been devoted to explore the fundamental links between LLTs and possible liquid-liquid critical points (LLCPs) in the supercooled liquid state(*3*). In the LLCP scenario predicted by several computer models of supercooled water, a first-order liquid-liquid transition occurs and the corresponding *P-T* curve ends up at a LLCP(*6–9*). The continuation of the curve beyond the LLCP is known as the Widom line, across which the LLTs become continuous and are smeared out if moving farther away from the LLCP(*10, 11*). Experimental and computational evidence of first- or higher-order LLTs and AATs has been reported in a wide range of materials, including water(*7, 12–14*), silicon(*15–19*), silica(*20, 21*), sulfur(*22*), tellurium(*23*), metallic glasses(*24, 25*), phase-change materials (PCMs)(*26–30*), and molecular liquids(*31, 32*). However, evidence of a LLCP remains scarce, with most findings coming from simulations, except for a few experimental cases(*22, 33*). Near a critical point (CP), materials exhibit extraordinary properties that of both fundamental and technological interest(*34–37*) .

Since glasses are commonly achieved by quenching liquids, the existence of AATs has long been considered as a manifestation of LLTs in the vitrified solid(*3*). However, the links between the two transitions are still elusive. A fundamental difference is that LLTs are thermodynamically driven transitions in a metastable or stable equilibrium (ergodic) state, while AATs are often mechanically induced by applying high pressure to a non-equilibrium (non-ergodic) state(*38*). The formulation of the free energy for a non-ergodic state is complicated by the effects of mechanical stress involved upon the pressure-induced transition(*38*). Hence, the comparison between LLTs and AATs is not straightforward(*3*). The existence and location of the LLCP have long been debated in relation to the LLT and AAT, particularly in supercooled water(*6, 13, 39–43*). In computational studies of water, the location of the transition lines in the pressure-temperature (*P-T*) diagram and their associated CP have been shown to depend on the form of the interaction potential(*7*). Some theories predict that the LLCP may even occur below the glass transition temperature $T_g$(*38*), though no evidence of such scenario has been found in experiments so far.

Recent studies suggested the existence of AAT and LLT in PCMs(*26, 27, 44*) such as $GeSb_2Te_4$, $Ge_{15}Sb_{85}$, GeTe, which are promising material candidates for non-volatile memory and neuromorphic computing devices(*45*). The AAT and LLT of PCMs are associated with remarkable property changes such as conductivity(*29, 46*) and fragilities(*27, 47, 48*), being discussed in the relevance to their functionalities in devices(*29, 49, 50*). While a temperature-induced LLT was reported for some PCMs in the supercooled liquid state by femtosecond diffraction experiments using X-ray free electron lasers(*27*), our recent study identified a pressure-induced AAT at room temperature in the prototypical PCM GeTe and in a related chalcogenide glass GeSe by a high-pressure synchrotron X-ray scattering experiment(*44*). The AATs are predominantly governed by the suppression of the structural distortion of local octahedral environments, namely the Peierls-like distortion (PLD), which is characterized by the periodic alternation of long and short bond sequences(*51*). The AAT proceeds by transforming the PLD into the undistorted octahedral-like coordination (OLC) without the long-and-short alteration, resulting in notable property changes (e.g. compressibility and electronic density of states) in the amorphous solids. The underlying mechanism responsible for the AAT in PCMs is found to be identical to that of the reported continuous LLTs in PCMs, where the PLD emerged in the supercooled liquid during rapid laser-induced melt-quenching processes(*27*). These findings highlighted the key role of two competing local structural motifs, the PLD and the OLC, in PCMs. However, it is not clear how the interplay of PLD and OLC would lead to a transition line in the *P-T* phase diagram that demarcates two disordered states with different locally favored structures.

In this work, we resolve the transition line for the amorphous pseudo-binary alloy $(GeSe)_{50}(GeTe)_{50}$ by identifying the pressure-induced AATs at temperatures below $T_g$. We show that this alloy exhibits a AAT driven by the same mechanism observed in its parent binaries GeSe and GeTe(*44*). Two *P-T* curves on the phase diagram are identified, representing the suppression of the PLD and the emergence of the OLC under compression. The separation of the two curves allows us to measure the "sharpness" of the AAT at each temperature, demonstrating an abrupt crossover from a continuous transition at lower temperatures to a "first-order-like" transition at 423 K. The crossover is characterized by the simultaneous occurrence of the suppression of PLD, the formation of OLC, and a maximum in the crystallization rate. The results can be rationalized in terms of a "two-state model" for the system based on the hierarchical model originally proposed for supercooled water, whose thermodynamic and dynamic anomalies are described by competing local bond-orientational ordering. The results for $(GeSe)_{50}(GeTe)_{50}$ present a striking analogy to the previous studies on the LLCP in supercooled water, where the transition between distinct local structural motifs coincide with the maxima in dynamic heterogeneity(*9*). We thus argue that the crossover hints at the existence of critical-like phenomena in the amorphous solid. The possibility of such behavior in a solid domain is intriguing, given that LLCP phenomena have been mostly explored only in liquid domains.

**Results**

### *In-situ* high-energy X-ray scattering under compression at various temperatures

We performed the *in-situ* high-pressure and elevated temperature X-ray scattering experiment at SPring-8 BL05XU (see the Supplementary Information). The experimental setup, combining a high-pressure instrument using the Paris-Edinburgh press, which has wide opening in horizontal plane for scattering experiment at high $2\theta$ angles, with high-energy X-ray beam (photon energy of 100.0646 keV), allows us to access a momentum transfer of $Q$ up to 27 Å$^{-1}$, which is almost twice as wide as that of a conventional setup, providing scattering data with high spatial resolution[52]. The pressure scans are performed isothermally for amorphous $(GeSe)_{50}(GeTe)_{50}$ at 300 K, 373 K, and 423 K below its glass transition temperature of 545 K[53]. Figure 1 shows the diffraction profiles $I(Q)$ at three temperatures. The positions of the 1$^{st}$ and the 2$^{nd}$ diffraction peaks are $Q_1$ = 2.01 Å$^{-1}$ and $Q_2$ = 3.44 Å$^{-1}$ at ambient pressure and 300 K. These values lie between those of amorphous GeSe ($Q_1$ = 2.11 Å$^{-1}$ and $Q_2$ = 3.59 Å$^{-1}$) and GeTe ($Q_1$ = 1.99 Å$^{-1}$ and $Q_2$ = 3.38 Å$^{-1}$)[44], as expected for a composition situated midway between the pseudo-binary parent compositions. The sample at 300 K remains fully amorphous up to 8.0 GPa (Figure 1a). Yet, at higher temperatures 373 K and at 423 K, it partially crystallizes during compression, as indicated by the emergence of Bragg peaks at high pressures (Figs.1b-c). The Rietveld refinement identifies the rhombohedral phase (space group: $R3m$) of crystalline $(GeSe)_{50}(GeTe)_{50}$ with only slight distortion from the NaCl-type cubic phase (space group: $Fm\bar{3}m$). Here we define the onset pressure of crystallization $P_{x\text{-}onset}$ as the lowest pressure, at which the intensity of Bragg reflections becomes sufficient for the Rietveld refinement to extract the lattice parameters. Thus, we obtain $P_{x\text{-}onset}$ = 3.8 ± 0.4 GPa for 373 K and $P_{x\text{-}onset}$ = 1.2 ± 0.3 GPa for 423 K (see the Figs. 1b-c).

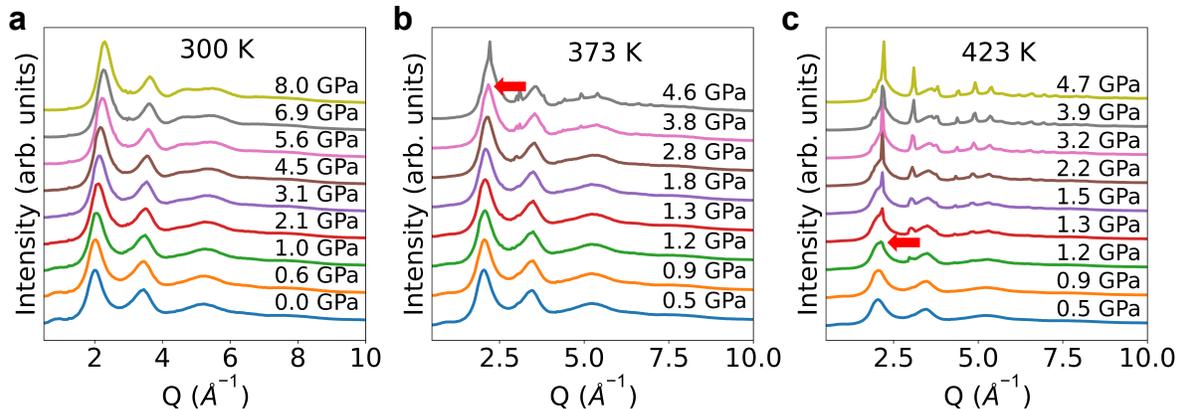

**Fig. 1. High-pressure X-ray scattering data $I(Q)$ of amorphous $(GeSe)_{50}(GeTe)_{50}$ at a)** 300 K, **b)** 373 K, and **c)** 423 K. Broad diffuse scattering peaks are observed at ambient pressure. The sample at 300 K remains fully amorphous up to 8.0 GPa. The red arrows show the locations of the <202> reflections of the rhombohedral phase to indicate the onset pressure of partial crystallization at 373 K and 423 K. We note that the tiny peaks around $Q \sim 2.9$ Å$^{-1}$ or 3.0 Å$^{-1}$ can be attributed to the components of the high-pressure cell such as BN (space group: $P6_3/mmc$) or MgO (space group: $Fm\bar{3}m$), which are not relevant to crystallization of the samples.

### The suppression of Peierls-like distortions

The PLD, the quasi-periodic long-and-short alternation in bond length, is a common structural feature of the low-pressure state of amorphous PCMs, while the undistorted OLC characterizes the high-pressure state[27, 44, 51, 54]. In the low-pressure state, the repetition of the long-and-short bond pairs of the PLD forms the medium-range orders. The length scale of such pairs defines the "period" of the repetition, which is nearly the doubled length of the short bonds of the OLC in real space. Therefore, the PLD manifests itself as a small "pre-peak" in $I(Q)$, and the position $Q_{ppk} \sim 1$ Å$^{-1}$ is nearly half of the position of the principal peak $Q_1 \sim 2$ Å$^{-1}$ in reciprocal space. Our recent study demonstrated that increasing pressure gradually diminishes the pre-peaks, ultimately causing them to vanish[44], reflecting the suppression of the PLD. Molecular dynamics simulation studies suggested that the suppression is mainly characterized by the compression of the long bonds[55]. We show later that the formation of the medium-range ordering of the OLC is represented by the emergence of a "high-order" diffraction peak in the high-pressure state around 4 Å$^{-1}$, being supported by the molecular dynamics simulation.

Figs. 2a-c show the log-scaled $I(Q)$ around 1.0 Å$^{-1}$ for amorphous (GeSe)$_{50}$(GeTe)$_{50}$. Under increasing pressure, the pre-peaks are reduced and eventually vanish in all three isothermal pressure scans. This indicates that pressure suppressed the PLD in the ternary system as well, suggesting an AAT similar to those observed in its parent binaries (amorphous GeSe and GeTe[44]). Figure 2d shows the pressure dependence of the integrated pre-peak intensity, normalized to its maximum value at the initial pressure for each isothermal scan, $i_{ppk} = I_{ppk}/I_{ppk}^{max}$. The data are also compared with those of amorphous GeSe and GeTe[44]. To characterize the pressure dependence, the data are fitted with a compressed exponential function

$$i_{ppk} = s_L * \exp(-(P/P_L))^{\gamma_L}, \tag{1}$$

where $s_L$ is a scaling factor, $P_L$ is a characteristic pressure for the suppression of PLD, and the exponent $\gamma_L$ is the shape parameter. A higher $\gamma_L$ indicates a more rapid decrease in intensity with increasing pressure.

Figure 2e shows the temperature dependence of $\gamma_L$. A significant increase from $\gamma_L = 1.2 \pm 0.1$ at 300 K to $\gamma_L = 5.5 \pm 0.5$ at 423 K occurs, indicating a more compressed exponential shape at 423 K, as also evidenced by the sharp drop in $i_{ppk}$ in Figure 2d (middle panel). Figure 2f shows the pressure derivative $-di_{ppk}/dP$, reflecting the suppression rate of the PLD. The narrow width and sharp maxima of $-di_{ppk}/dP$, at 423 K, along with the high $\gamma_L$ from the fits, indicates a significantly faster suppression rate of the PLD comparing to those at 300 K and at 373 K. This suggests that at 423 K, the AAT becomes markedly sharper, behaving as a first-order-like transition. The characteristic pressures $P_L$ are $1.3 \pm 0.1$ GPa at 300 K, $1.6 \pm 0.1$ GPa at 373 K, and $1.37 \pm 0.01$ GPa at 423 K. $P_L$ at 423 K is 14 % lower than that at 373 K, suggesting a negative slope of the AAT line on the $P$-$T$ phase diagram (see also the Supplementary Info for details).

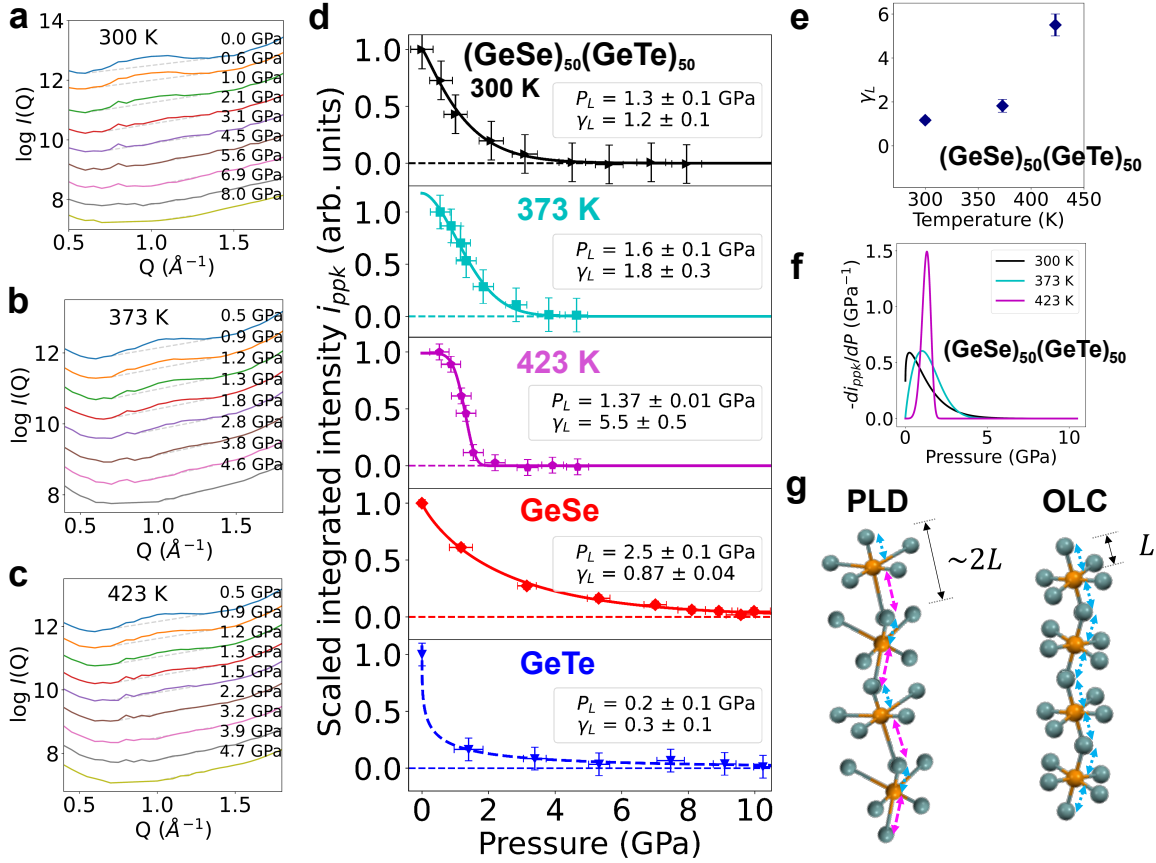

**Fig. 2. The suppression of Peierls-like distortions.** The pressure dependence of the pre-peak of *I(Q)* in the low-*Q* range at **a)** 300 K, **b)** 373 K, and **c)** 423 K for (GeSe)$_{50}$(GeTe)$_{50}$. The gray dashed lines are the baseline by linear interpolation. Note that the tiny peaks around $Q \sim 0.8$ Å$^{-1}$ are due to the background scattering from the sample cells and should not be confused with the pre-peak. **d)** The pressure dependence of $i_{ppk}$ of (GeSe)$_{50}$(GeTe)$_{50}$ at 300 K (black), 373 K (cyan), and 423 K (magenta), as well as that of GeSe(*44*) (red), and GeTe(*44*) (blue) at 300 K. The data points are fitted by Equation (1). **e)** The temperature dependence of $\gamma_L$ suggests that the AAT becomes sharper at higher temperatures. **f)** The derivative $-di_{ppk}/dP$ represents the suppression rate of the PLD. The sharp peak at 423 K demonstrates a sharp transition, while the transitions at other temperatures are smeared out. **g)** The schematics of the periodic-like array of long and short bonds of the PLD, and that of the OLC if all long bonds have been compressed to short bonds. The AAT is associated not only with a halved periodic length ($2L$ to $L$) but also with doubling of the repetition number (e.g., 4 to 8, as illustrated).

### The emerging of local octahedral-like coordinate motifs

Since the PLD only represents the low-pressure state before the AAT, demonstrating the change in AAT characters requires a quantitative evaluation of the emergence of the OLC in the high-pressure state. In general, the intensity of a diffraction peak is sensitive to the number of periodic repetitions of a structural motif(*56*). Figure 2g schematically shows a quasi-periodic array of alternating long and short bonds of the PLD, and that of the OLC if all long bonds have been compressed to short bonds. Assuming that an array contains the same number of atoms before and after the AAT, the AAT is associated not only with a halved periodic length ($2L$ to $L$), but also with a doubling of the number of repetitions (e.g. 4 to 8, as illustrated). This substantial increase in the number of repetitions will lead to the intensity growth of diffraction peaks related to the OLC in the entire $Q$ space. Since the OLC is not dominant in the low-pressure state, their associated peaks are vanishingly small; however, as pressure transforms the PLD into the OLC, we expect the growth of the OLC-related peaks, or emergence of a "new" peak in $I(Q)$.

Figure 3a shows the medium $Q$-range of the $I(Q)$ from 3.0 to 9.0 Å$^{-1}$ for $(GeSe)_{50}(GeTe)_{50}$ at 300 K. The profile shape of the $I(Q)$ around $Q \sim 4.5$ Å$^{-1}$ varies from a monomodal to bimodal upon compression, suggesting the emergence of a new periodic-like medium-range order in the structure. We find the same behavior in the profile shapes at 373 K and 423 K for $(GeSe)_{50}(GeTe)_{50}$, as well as for GeSe, and GeTe (Figure S8). This suggests that this medium-range structural order exists both in the binary and the pseudo-binary compositions.

To identify the structural origin, we decompose the $I(Q)$ into individual profile functions by applying the profile fitting analysis (see the Supplementary Info for the analytical conditions, fitting results and refined parameters). For convenience, we index the profile functions as the 1$^{st}$, 2$^{nd}$, 3$^{rd}$ and so on in an increasing order of the $Q$-positions. The 3$^{rd}$ peak is responsible for the emerging feature, as shown in the inset of Figure 3a (for 300 K) and in Figure 3b (for 423 K), largely overlapping with the neighboring 4$^{th}$ peak.

We deliberately remove the 3$^{rd}$ peak from the $I(Q)$ and Fourier transform the resulting scattering profile to obtain the reduced pair distribution function (PDF) $G(r)$. Figure 3c shows the comparison between the $G(r)$ from the $I(Q)$ without the 3$^{rd}$ peak and with the one from the original $I(Q)$. A pronounced change is the shift of the 1$^{st}$ peak position $r_1$ of $G(r)$ to the low-$r$ side after removing the 3$^{rd}$ peak of $I(Q)$, while the positions of the 2$^{nd}$ peaks $r_2$ of $G(r)$ keep nearly identical. Such difference is qualitatively consistent with the peak shifts introduced by the formation of the PLD, where splitting and unequal occupation of the first coordination shell effectively lowers $r_1$ with respect to $r_2$, increasing the ratio $r_2/r_1$[27]. The differences are also found in the height of the 2$^{nd}$ peak of $G(r)$ and in the depth of the valley between the 1$^{st}$ peak and the 2$^{nd}$ peak of $G(r)$, while the two $G(r)$'s are nearly identical in higher $r$-range. Such a modulation suggests that the 3$^{rd}$ peak of $I(Q)$ reflects the redistribution of atoms from the second shell to the first shell in real space, which is consistent with the suppression of the long bonds of the PLD that extend to the second shell[55]. The results suggest that the emergence of the 3$^{rd}$ peak is a footprint of the formation of the OLC, which will be confirmed by our molecular dynamic simulations, as shown in the next section.

In this context, we can quantify the OLC by measuring the intensity of the 3$^{rd}$ peak of $I(Q)$ normalized to the sum of the intensities of the 3$^{rd}$ and 4$^{th}$ peaks, denoted as $i_3 = I_3/(I_3 + I_4)$, where $I_3$ and $I_4$ are the integrated intensities of the 3$^{rd}$ peak and 4$^{th}$ peak, respectively. Figure 3d shows the pressure dependence of $i_3$ in the pseudo-binary and binary systems. We fit the data points of $i_3$ with a modified compressed exponential function,

$$i_3 = c - s_H * \exp((-P/P_H)^{\gamma_H}) \tag{2}$$

where $c$ is a constant, $s_H$ is the scaling factor, the exponent $\gamma_H$ is the shape parameter, and $P_H$ is the characteristic pressure for the formation of the OLC. We find that $i_3$ reaches a plateau of 0.4 at 3 GPa in $(GeSe)_{50}(GeTe)_{50}$ at 423 K. At room temperature, the plateau for GeSe is 0.41 at $\sim$9 GPa, while the one for GeTe is 0.39 at $\sim$ 4 GPa. Thus, the plateau values of $i_3$ are consistently found around 0.4, suggesting the complete transformation of the structural motifs from the PLD to the OLC. The $P_H$ and $\gamma_H$ of $(GeSe)_{50}(GeTe)_{50}$ at 300 K and 373 K are estimated by assuming that $i_3$ reaches the same plateau value of 0.4 as for the case of 423 K. The resulting exponent $\gamma_H$ increases from $1.6 \pm 0.3$ at 300 K to $4.6 \pm 0.7$ at 423 K, indicating a significant sharpening of the AAT. This exhibits remarkable consistency with the temperature dependence of $\gamma_L$. In Figure 3b, we highlight the pressure response of the 3$^{rd}$ peak and the 4$^{th}$ peak at 423 K. We observe not only the rapid growth of the 3$^{rd}$ peak but also the pronounced sharpening of the 4$^{th}$ peak (see also Fig. S9 for comparison with the other temperatures).

Figure 3e shows the pressure derivative $-di_3/dP$, representing the rate of formation of the OLC at each temperature. The narrow width and high sharpness of the $-di_3/dP$ peak at 423 K is consistent with the rapid suppression rate of the PLD observed in Figure 2f at the same temperature, supporting the notion that AAT is a first-order-like transition at 423 K. The characteristic pressure $P_H$ decreases significantly from $4.7 \pm 0.5$ GPa at 300 K to $1.8 \pm 0.1$ GPa at 423 K, which is again consistent with the lowering $P_L$ at elevated temperature and suggests a negative slope of the AAT line on the $P$-$T$ phase diagram.

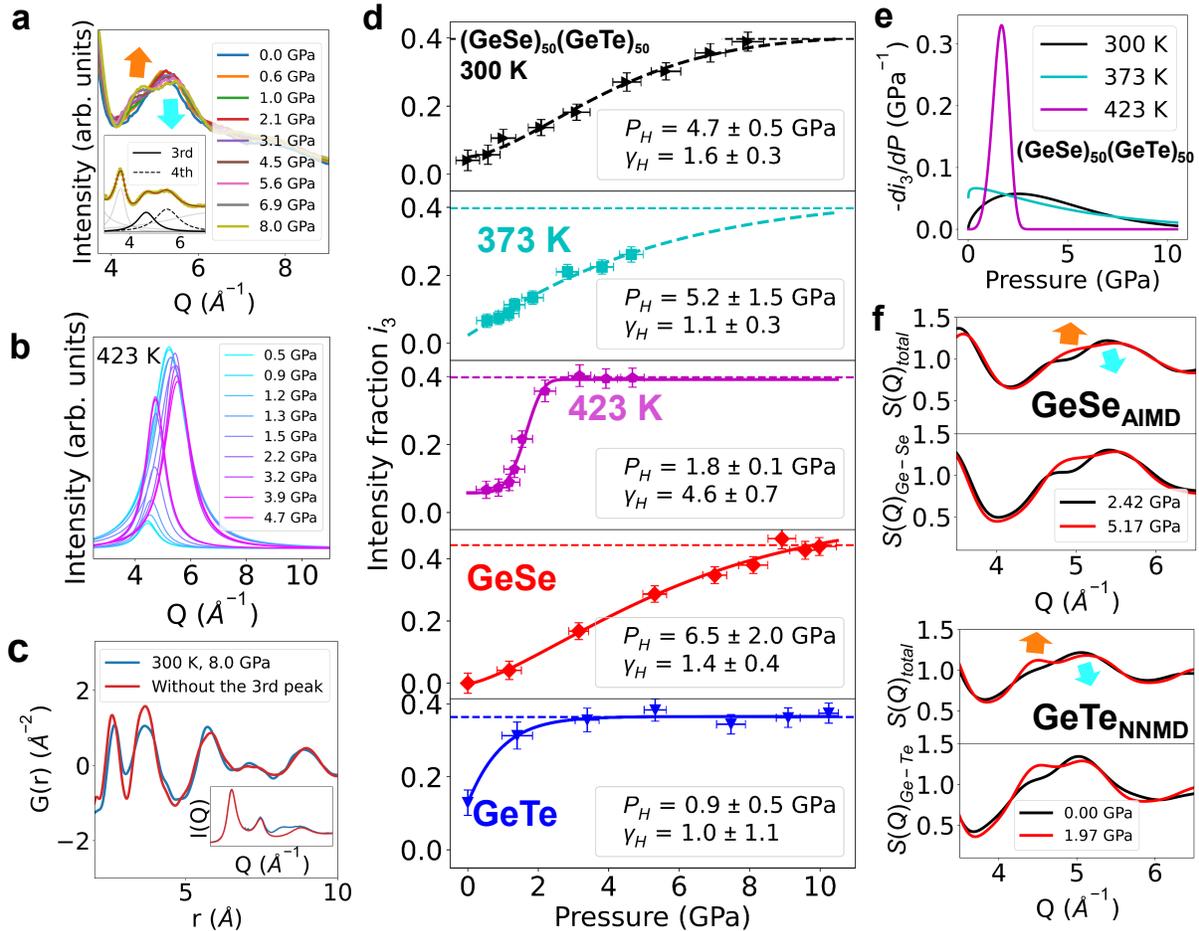

**Fig. 3. The emergence of the octahedral-like coordination motifs. a)** The pressure response of the $I(Q)$ at 300 K for (GeSe)$_{50}$(GeTe)$_{50}$. A peak-like feature emerges around $Q \sim 4.5$ Å$^{-1}$, modulating the shape of $I(Q)$ from monomodal to bimodal. The inset is the representative result of the profile fitting, where the 3$^{rd}$ peak is responsible for the emerging feature. **b)** The fitted profiles of the 3$^{rd}$ peak and 4$^{th}$ peak of $I(Q)$ for (GeSe)$_{50}$(GeTe)$_{50}$ at 423 K at different pressures. **c)** The reduced pair distribution functions $G(r)$ converted from the original $I(Q)$ with (blue) and without the 3$^{rd}$ peak (red). The inset shows the $I(Q)$ with (blue) and without (red) the 3$^{rd}$ peak. **d)** The pressure dependence of the intensity fraction $i_3$ for (GeSe)$_{50}$(GeTe)$_{50}$ at 300 K (black), 373 K (cyan), and 423 K (magenta), as well as for GeSe(44) (red), and GeTe(44) (blue) at 300 K. The data points are fitted by Equation (2). The $i_3$ at 423 K is characterized by a smaller $P_H$ and a rapid rise with increasing pressure (i.e. higher $\gamma_H$), compared to its gradual increase at lower temperatures. **e)** The derivative $-di_3/dP$, representing the formation rate of the OLC, shows a substantially sharper peak at 423 K than that at 300 K and 373 K. **f)** The simulated $S(Q)$ of amorphous GeTe and GeSe under various pressures obtained by the NNMD and AIMD simulations, respectively, reproducing the monomodal-to-bimodal feature as observed in experiments.

**Molecular dynamics simulations**

To support the interpretation of the 3rd peak of the $I(Q)$, we performed molecular dynamics simulations for the amorphous state of the parent compositions GeTe and GeSe using a machine-learned potential and an AIMD approach, respectively. The models were generated by a standard melt-quench approach with a quenching rate of $10^{13}$ K s$^{-1}$. Figure 3f shows the calculated total structure factors $S_{total}(Q)$ of amorphous GeTe at two pressures, which reproduce the monomodal-to-bimodal shift of the profile shape of experimental S($Q$) around $Q \sim 4.5$ Å$^{-1}$. The calculated $S_{total}(Q)$ for amorphous GeSe shows a similar rise at higher pressure, around $Q \sim 4.9$ Å$^{-1}$. The results agree well with experimental observations in both binary alloys and are also consistent with the observed monomodal-to-bimodal shift in the pseudo-binary composition (GeSe)$_{50}$(GeTe)$_{50}$. The calculated partial structure factors $S_{Ge-Te}(Q)$ and $S_{Ge-Se}(Q)$ also show the monomodal-to-bimodal shifts in their profile shapes (Figs. 3f and S10). The strong correlation between $S_{total}(Q)$ and $S_{Ge-Se(Te)}(Q)$ suggests that the periodic-like structure of these heteropolar bonds are responsible for the emergence of the 3rd peak.

It is well known that AIMD models of amorphous GeTe and GeSe contain some tetrahedral structures, which are promoted by the presence of homopolar Ge-Ge bonds[51, 57, 58]. Our previous study demonstrated that pressure transforms the PLD into the OLC, and meanwhile reduces the fraction of tetrahedral motifs (e.g. 36.3 % to 21.5 % in GeTe from 0.0 GPa to 1.97 GPa)[44]. Thus, the emergence of the 3rd peak of $I(Q)$ cannot be attributed to tetrahedral motifs. On the contrary, the OLC structural motif is consistent with the periodic-like structure of the heteropolar bonds (Figure 2g). Moreover, as mentioned above, the emergence of a peak-like feature at high-$Q$ is consistent with the larger repetition number of a structural motif of the OLC with respect to that of the PLD (Figure 2g). Thus, only the formation of the periodic-like structure of the OLC can account for the emergence of the 3rd peak. Further details from the MD simulations are provided in Figs. S10 and S11.

**The crossover from the continuous to first-order-like AATs**

The remarkable difference between $P_L$ and $P_H$ at a given temperature shows that the suppression of PLD and the emergence of OLC do not necessarily occur simultaneously. This indicates that the sharpness of the AATs varies, depending on temperature. It allows us to define the difference between the characteristic pressures $\Delta P = P_H - P_L$ as a quantitative measure of the sharpness of the AAT. $\Delta P$ is $3.4 \pm 0.5$ GPa at 300 K and $3.6 \pm 1.5$ GPa at 373 K. However, $\Delta P$ decreases substantially to $0.4 \pm 0.1$ GPa at 423 K. The changes in sharpness indicate a crossover from continuous transitions at 300 K and 373 K to a first-order-like (discontinuous) transition at 423 K. This trend is also reflected in the $\gamma_L$ and $\gamma_H$ values, where the rate of suppression of PLD and of formation of OLC reaches the highest value at 423 K. This crossover is reminiscent of the phase diagrams generated from two-state models, which we will take up in Discussion.

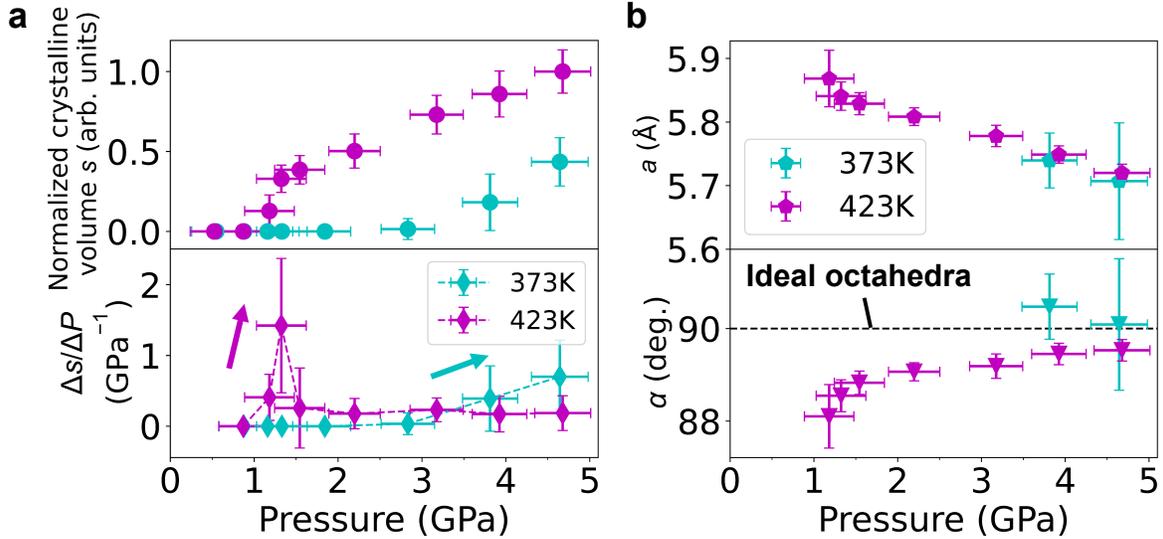

**Fig. 4. The crystallization to the rhombohedral (GeSe)$_{50}$(GeTe)$_{50}$. a)** The pressure dependence of the normalized crystalline volume *s*. The pressure derivative $\Delta s/\Delta P$ indicates the crystallization rate with respect to pressure. A remarkable maximum of $\Delta s/\Delta P$ is observed at 1.3 GPa and 423 K. **b)** The pressure dependence of the lattice parameters *a* and *α*. The horizontal dashed line in the bottom plot shows the cell angle $\alpha = 90°$ of the cubic phase with the ideal octahedral coordination.

**Crystallization behavior related to PLD- and OLC-dominated amorphous phases**

Given that crystals and glasses essentially possess the same interaction potential, their (local) structural responses to pressure tends to be similar(*3, 16*). The rhombohedral crystals of (GeSe)$_{50}$(GeTe)$_{50}$ show the alternation of the long-and-short bonds, called the Peierls-distortion (PD) -- the crystalline counterpart of the PLD. Earlier studies showed that the vanishing of the pre-peak and the crystallization of the amorphous state occur in a pressure range, where crystalline GeTe undergoes a pressure-induced rhombohedral-to-cubic transition (i.e. suppression of the PD)(*44, 59*). Given that the amorphous and crystalline states are closely interconnected, we also investigate the crystallization behavior in relation to the AAT.

Figure 4a shows the pressure dependence of the crystalline volume normalized to its maximum volume at 423 K. The crystalline volume fraction is determined by profile fitting of total scattering intensities. Specifically, the Rietveld refinement of the Bragg peaks for the crystalline phases yields a scaling factor, *s*, which is approximately proportional to the crystalline volume within the sample's scattered volume (see the Supplementary Information). The value of *s* shows a sharp rise from 1.2 to 1.3 GPa at 423 K, followed by a stable linear trend, resulting in a maximum and the subsequent plateau in the pressure derivative $\Delta s/\Delta P$ (the bottom plot of Figure 4a). A peak in $\Delta s/\Delta P$ at 1.3 GPa indicates a sharp maximum of the crystallization rate with respect to pressure; yet, no such maximum is observed at lower temperatures. Such an anomaly cannot be explained by the classical nucleation and growth theories, as will be further discussed in the next section.

Figure 4b shows the pressure dependence of the lattice parameters (i.e., lattice constant *a* and cell angle *α*) of rhombohedral (GeSe)$_{50}$(GeTe)$_{50}$. At 373 K, the *α* of the rhombohedral unit cell is 90° within error above 3.8 GPa, while at 423 K the value of α systematically increases from 88.1(7)° at 1.2 GPa to 89.5(2)° at 4.6 GPa. These angles are close to 90°, indicating that the crystallized phase is nearly the cubic phase with ideal octahedral coordination.

## Discussion

### Locally favored structures and the two-state model

As pointed out by Angell and others, liquid and amorphous PCMs exhibit thermodynamic and dynamic anomalies, resembling those of supercooled water, including anomalies in density(*50*, *60*), heat capacity(*60*, *61*), and isothermal compressibility(*23*, *60*), LLT(*27*), and dynamical crossover (i.e. fragile-to-strong transition)(*47*). Water-like anomalies have been proposed to be understood in the hierarchical two-state models, where the locally favored structures (LFS) are considered as the relevant order parameter. For instance, tetrahedral coordination is the energetically favored in water, Si(*15*) and Ge(*62*), whereas icosahedral packing is entropically favored in some metallic liquids(*63*). In the case of water, the interplay between two local structural motifs—the tetrahedral structure and the disordered normal liquid structure—provides a unified description of water's anomalies(*64*). In the present system, the AATs are characterized by the transformation between the PLD and OLC motifs. Based on these analogies, a water-like two-state model can be applied to the present system to describe the AATs and the crystallization behavior in the *P-T* phase diagram (see Figure 5). We note that the non-ergodicity of amorphous phases may render the AAT different from the thermodynamically driven LLT(*3*). Nonetheless, we draw a phenomenological parallel to the LFS and its *P-T* responses of supercooled water.

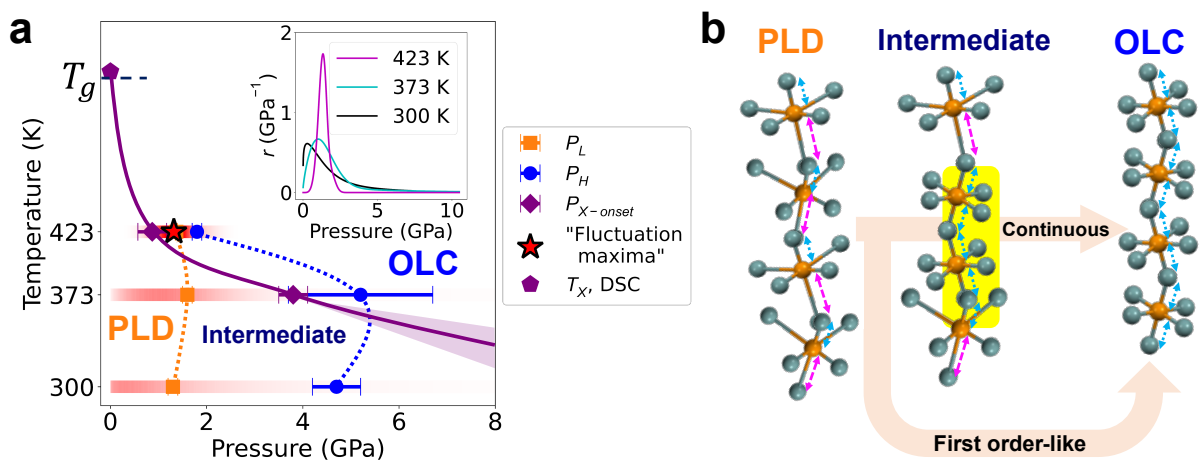

**Fig. 5. The crossover from continuous to first-order-like AATs. a)** *P-T* phase diagram of amorphous $(GeSe)_{50}(GeTe)_{50}$ based on the local structural motifs. The orange line connects the characteristic pressures of the suppression of the PLD ($P_L$), while the blue lines connect those of the emergence of the OLC ($P_H$). The two lines divide the diagram into three regions: the PLD-favored region, the OLC-favored region, and the intermediate region. The purple curve connects the onset pressure of crystallization ($P_{X-onset}$) with the crystallization temperature $T_X$ from the differential scanning calorimetry (DSC) measurement(*53*). The red star shows the *P-T* point of the $\Delta s/\Delta P$ maximum, labelled as fluctuation maxima. The inset shows $r$ to indicate the sharpness of the transitions from the PLD to the OLC. The $r$ values are also shown in the color contour in the phase diagram, with a red gradient for each temperature. In contrast to 423 K, it spans a much wider pressure range at 300 and 373 K. **b)** The schematic drawing of the pressure-induced local structural changes. The period $L$ of the OLC is approximately half of that of the PLD. The intermediate state is characterized by the partially shortened long bonds of the PLD and the breakdown of the medium-range long-and-short bond alteration (i.e. pre-peak vanishing). Despite the formation of some local chains of short bonds (i.e. growth of the 3rd peak), a considerable number of distorted short-range octahedral motifs continue to coexist.

**Water-like crossover in phase-change materials**

As described in the Results, a significant reduction in the sharpness $\Delta P$ from $\Delta P = 3.4 \pm 0.5$ GPa at 300 K to $\Delta P = 0.4 \pm 0.1$ GPa at 423 K, suggests a crossover of the AAT from a continuous transition to a first-order-like transition. Figure 5a shows the characteristic pressures $P_L$ and $P_H$ on the $P$-$T$ phase diagram of amorphous $(GeSe)_{50}(GeTe)_{50}$. The curve connecting the $P_L$ values (i.e. $P_L$-curve) shows the boundary of the region, where the PLD is favored as local structure, while the other curve connecting the $P_H$ values (i.e. $P_H$-curve) demarcates the region where the OLC is favored. The domain between the two curves is regarded as the "intermediate" range, where neither of the two motifs are dominant. Figure 5b shows the schematics of the local structures. The intermediate state can be represented by a partially-suppressed PLD. Specifically, the local suppression of the long bonds induces the breakdown of the medium-range long-and-short bond alternation (i.e. the vanishing of the pre-peak). Depending on the temperature, the AAT may follow two distinct pathways: 1) a continuous transition involving the intermediate state characterized by partially suppressed long bonds of the PLD and the gradual formation of local chains of the short bonds (i.e. the emergence of the $3^{rd}$ peak of $I(Q)$); 2) a first-order-like (discontinuous) transition to the OLC. Therefore, the AAT at 423 K represents a discontinuous transition from the PLD to the OLC, while at lower temperatures the AAT changes continuously between the two structural motifs. The two distinct transition behaviors can also be seen in the inset of Fig.5a. We use the sum of both pressure derivatives, $r = (-di_{ppk}/dP) + (-di_3/dP)$, as a measure of the effective rate of overall structural change from the PLD to the OLC. Clearly, the remarkable sharpness of the $r$ peak at 423 K indicates a first-order-like transition, while the smeared-out peaks at 300 and 373 K indicate higher-order continuous transitions, supporting the crossover of the AAT. We note that the transition from PLD to OLC is essentially a restoring of the local broken symmetry.

Such crossover from continuous to first-order like transition in the local structure is an important feature of the LLT of supercooled water around the LLCP in computer models[64]. The LLCP was first reported in the ST2 model of water[39]. Recent studies provided evidence for its existence in two more realistic water models, TIP4P/2005 and TIP4P/Ice[6]. The hierarchical two-state description of TIP4P/2005 shows the drastic sharpening in the temperature-induced conversion from the tetrahedral structure to the disordered structure when the isobaric condition is modified from 0.1 MPa to the critical pressure $P_c \sim 180$ MPa[64]. Their results are remarkably analogous to our observation in the pressure response of the pre-peak and the $3^{rd}$ peak (e.g. rise of the shape parameter $\gamma_L$ and $\gamma_H$ at 423 K). The present system allows us to trace the pressure-induced structural transformation by measuring both $P_L$ and $P_H$, and the sharpness $\Delta P$, experimentally reproducing the features of water-like crossovers predicted so far only in computer models.

**The case for a critical-like point in the amorphous solids**

In Fig. 5a, we also plot the curve connecting the onset pressure of crystallization $P_{X-onset}$ on the phase diagram. The curve shows a convex trajectory, deviating both from the $P_L$-curve and the $P_H$-curve. This deviation is counter-intuitive, since one would expect that crystallization would set in when the local OLC structure in the amorphous phase matches the octahedral coordination of the NaCl-type crystal structure, lowering the interfacial energy and the barrier of nucleation. However, the observed discrepancy indicates that static structural similarities are not the only factor facilitating crystallization.

In addition, the pressure derivative of the normalized crystalline volume $\Delta s/\Delta P$ (Fig. 4a lower panel) at 423 K exhibits a maximum at 1.3 GPa, reflecting that the crystallization rate reaches a maximum value at this pressure. Yet, the static local structures (i.e. suppression of PLD and emerging of OLC) exhibit only monotonic changes with pressures. A plausible explanation is that dynamics fluctuations may play an important role in the observed crystallization behavior, as atomic mobility is the limiting factor for both nucleation and growth rates of crystallization. In other words, $\Delta s/\Delta P$ maximum at 423 K and 1.3 GPa implies a maximum in dynamical fluctuations (i.e. dynamic heterogeneity) at this specific $P$-$T$ combination. In such scenario, the broad distribution of the relaxation times for atomic dynamics has a long tail extending to fast timescales in local regions, facilitating the crystallization. As shown in Fig.5a, the crystallization rate maximum overlaps with the characteristic pressure $P_L = 1.37 \pm 0.01$ GPa, which is the locus of the abrupt crossover of the AAT from a continuous to the first-order-like transition. This is reminiscent of the critical-like phenomena near the LLCP in water-like model systems, where the crossover from continuous to first-order transitions coincides with the maximum in dynamical fluctuations(9). In their hierarchical two-state model of water, Shi and Tanaka argued that the maxima of dynamical fluctuations, defining the "dynamic Schottky line" in the $P$-$T$ diagram, often diverge from the maxima of density and/or entropy fluctuations (i.e. static Schottky line); however, these two lines converge at the LLCP in the model liquids(9). The relationship between a metastable critical point and the crystallization rate has also been explored in computational studies of protein crystallization, where the presence of a critical point strongly reduces the free-energy barrier, resulting in orders-of-magnitude increase in the crystal nucleation rate(65).

The phenomenological analogy to the LLCP in water-like model systems suggests the possibility that a critical-like point may exist at the crossover point around 1.3 GPa and 423 K in the amorphous solid state of the present system. The scenario of a critical point in amorphous solid is in stark contrast to the speculated LLCP commonly discussed in the supercooled liquid state in most computer models(7). According to the two-state model(38), the location of the critical point on the phase diagram depends on the energy differences between the two local structural motifs and the strength of their cooperative formation(3, 38). Notably, the model suggests that the critical point would be located well below $T_g$ when cooperativity is negligibly weak(3).

The crossover from continuous to first-order-like AATs in $(GeSe)_{50}(GeTe)_{50}$ presents a striking parallel to the LLCP of supercooled water. Water-like thermodynamic anomalies have been discovered in elemental liquid Te(66), $Ge_{15}Te_{85}(67)$, Se-Te(68), and other compositions based on group-IV, V, VI(61, 69). Tuning compositions can be considered equivalent to tuning interaction potentials, resulting in various anomalous behaviors as predicted by theoretical models(70). In this context, the location of the critical-like point and the sharpness of the AAT might be adjusted by systematically altering alloy components, for instance, in the pseudo-binary $(GeSe)_x(GeTe)_{1-x}$ alloys. The GeSe as a binary alloy, with strong covalent bonding, exhibits a significantly low sharpness of $\Delta P = 4.0 \pm 2.0$ GPa, as compared to GeTe, which is characterized by weaker covalent bonding and has a pronounced sharpness of $\Delta P = 0.4 \pm 0.1$ GPa. It was reported that elemental Ge undergoes a first-order liquid-liquid transition, while Te shows a gradual continuous transition below its melting temperature at ambient pressure. The existence of a critical-like point in amorphous solids would lead to remarkable property changes such as critical-like fluctuations, irrespective of material structural and chemical details. Near such a point, phase changes might be drastically accelerated due to fluctuations(64, 65) and the optical properties might be affected by critical opalescence(71). Further studies may test this hypothesis, for example, by measuring order parameter fluctuations and testing the power law scaling as predicted by Wilson's renormalization group theory. Identifying critical-like phenomena in amorphous solids would also provide a compelling parallel to the critical phenomena seen in crystalline solids such as the order-disorder transition in $Fe_{50}Co_{50}(72)$. Mechanically-induced critical phenomena for amorphous solids has been suggested in the yielding transition in computer models(73). Lastly, we note that owing to the non-equilibrium nature of amorphous solids, aging may alter local structures, possibly introducing subtle effects in AAT behaviors. Earlier studies of PCMs suggested that aging reinforces PLD. Thus, one might speculate a higher pressure required to complete the AAT, which is of interest for future studies.

**Acknowledgements**

The synchrotron radiation experiments were performed at BL05XU of SPring-8 with the approval of the Japan Synchrotron Radiation Research Institute (JASRI) (Proposal No. 2023B1387). SW is grateful to YD Cheng, J Pries, M Wuttig for help with sample preparation. **Funding:** This work was supported by a research grant (42116) from VILLUM FONDEN (S.W.). We thank the Danish Agency for Science, Technology, and Innovation for funding the instrument center DanScatt. This work was financially supported by the Japan Society for the Promotion of Science (JSPS) KAKENHI Grant No. 20H00201 (Y.K.), 24H00415, 21H05235 (E.N.). **Authors contributions:** T.F., R.M., and S.W. jointly supervised this work. T.F., S.W., E.N., and Y.K. initiated the project. T.F., Y.K., J.M., S.T., A.M., S.K., K.O., I.I., Y.H., M.Y, and E.N. contributed to the experimental methods and performed the experiment with samples from S.W.. T.F. and S.W. analyzed experimental data. Y.C., R.M., D.C., and M.B. developed computational methods and performed AIMD and NNMD simulations. T.F., S.W., and R.M. wrote the manuscript with important input from Y.K., M.B., and others. **Competing interests:** The authors declare no competing interests. **Data and materials availability:** All data is available in the manuscript and the Supplementary Information.

**Supplementary Materials**

Materials and Methods

Tables S1 and S2

Figs. S1 to S11

# Supplementary Information:

# From Continuous to First-Order-Like: Amorphous-to-Amorphous Transition in Phase-Change Materials

Tomoki Fujita[1]*, Yoshio Kono[2], Yuhan Chen[3], Jens Moesgaard[1], Seiya Takahashi[4], Arune Makareviciute[1], Sho Kakizawa[5], Davide Campi[6], Marco Bernasconi[6], Koji Ohara[7], Ichiro Inoue[8], Yujiro Hayashi[8], Makina Yabashi[8], Eiji Nishibori[4], Riccardo Mazzarello[3]*, Shuai Wei[1,9]*

[1]*Department of Chemistry, Aarhus University, 8000 Aarhus C, Denmark.*
[2]*Department of Physics and Astronomy, Kwansei Gakuin University, Sanda 669-1330, Japan.*
[3]*Department of Physics, Sapienza University of Rome, Rome 00185, Italy.*
[4]*Department of Physics, Faculty of Pure and Applied Sciences and Tsukuba Research Center for Energy Materials Science (TREMS), University of Tsukuba, Ibaraki 305-8571, Japan.*
[5]*Japan Synchrotron Radiation Research Institute, 1-1-1 Kouto, Sayo-cho, Sayo-gun, Hyogo 679-5198, Japan.*
[6]*Department of Materials Science, University of Milano-Bicocca, I-20125 Milano, Italy.*
[7]*Faculty of Materials for Energy, Shimane University, Matsue, Shimane 690-8504, Japan.*
[8]*RIKEN SPring-8 Center, 1-1-1 Kouto, Sayo-cho, Sayo-gun, Hyogo 679-5148, Japan.*
[9]*iMAT Centre for Integrated Materials Research, Aarhus University, Denmark.*

## Materials and Methods

### Samples and *in-situ* high-pressure and elevated temperature experiment

Amorphous $(GeSe)_{50}(GeTe)_{50}$ samples were prepared by direct current magnetron sputtering deposition (base pressure ~$10^{-6}$ mbar and argon flow of 20 sccm) with co-sputtering mode from binary stoichiometric targets with purity higher than 99.99%. XRF indicates that the as-deposited composition is slightly on Te-rich side with ~5% more Te. As-deposited amorphous layers with a thickness of 5-8 um were scraped off from substrates and cut into small flakes for X-ray scattering experiments. We performed the *in-situ* high-pressure and elevated temperature X-ray scattering experiment at the BL05XU beamline of SPring-8($52$). A high-flux pink beam at the photon energy of an X-ray beam of 100.0646 keV, which was generated with a double multilayer monochromator at the BL05XU($74$). The high-flux pink beam allows us to efficiently measure the $I(Q)$ with a good signal-to-noise ratio, providing the data in total 26 combinations of the $P$-$T$ points in 6 shifts (48 hours). The variation in exposure time and measurement interval was below 5 % (see Table S2). We conducted three pressure scans at 300 K, 373 K, and 423 K. The pressure ranges were from an ambient condition to 8.0 GPa at 300 K, 0.5 GPa to 4.6 GPa at 373 K, and 0.5 GPa to 4.7 GPa at 423 K. We also measured one data point at an ambient condition after pressure release. The two types of high-pressure cells were employed: the cupped-Drickamer-Toroidal (CDT)-type anvil cell for room temperature measurements and the high-temperature cell for the elevated temperature measurements($75$). A graphite heater and a thermocouple were incorporated in a high-temperature cell, and sample temperature were monitored throughout the measurement. A pressure standard was placed along with sample, which were a gold thin film for room temperature measurements and a MgO sleeve for elevated temperature measurements. Pressure was calibrated by the volume ratio of the unit cell of the pressure standards, using the equations of state of Au($76$) and MgO($77$). We loaded the sample cells in the Paris-Edinburgh press, and the diffraction data were collected by scanning $2\theta$ angle by using two-point detectors up to 31.8 degrees, corresponding to $Q = 27$ Å$^{-1}$.

### Rietveld refinement

The procedure of data processing was described in Fujita *et al.* (2023)($44$). We performed the Rietveld refinement on the program Synchrotron Powder($78$). The analytical range was from 0.6° to 31.7° in $2\theta$. The pseudo-Voigt function was employed to express the Bragg peaks. The rhombohedral phase of $(GeSe)_{50}(GeTe)_{50}$ (space group: $R3m$), representing the same structure as the rhombohedral GeTe with 50 % of Se occupancy in Te sites, provided the best description for all the data with partial crystallization. The Bragg peaks observed after pressure release were characterized by the same rhombohedral phase, with no signs of phase separation. The scaling factor, lattice parameters, and peak width parameters are refined. The $I(Q)$ are normalized to the incident flux prior to the refinement and no significant changes of the fractional coordinates are identified. Therefore, the refined scaling factor is approximately proportional to the illuminated volume of crystalline sample($56$). The pressure dependencies of the refined parameters (i.e. scaling factors and lattice parameters) are shown in Figure 4 in the main text. Figs. S2 and S3 show the fitting results of Rietveld refinement of $(GeSe)_{50}(GeTe)_{50}$ at 373 K and at 423 K. Notably, tiny "humps" around $2\theta \sim 3.5°$ can also be attributed to the background scattering from the high-pressure cell components such as 100 and 101 reflections of the hexagonal BN (space group: $P6_3/mmc$) or 200 reflection of the cubic MgO (space group: $Fm\overline{3}m$). 024 and 220 reflections of the rhombohedral $(GeSe)_{50}(GeTe)_{50}$ could superpose with these background scattering, which makes it difficult to distinguish them at early stages of crystallization. Therefore, we defined the onset pressure of crystallization as the minimum pressure for each temperature scan, where the Bragg peaks are sufficiently pronounced, and the lattice parameters are not diverged during the Rietveld refinement.

**Profile fitting**

We performed profile fitting simultaneously with Rietveld refinement to extract amorphous diffraction profiles from $I(Q)$. We employed the split-type Pearson-VII function(*79*) to express the asymmetric shape of amorphous diffraction peaks. The amorphous scattering was expressed as the sum of 10 profile functions. We indexed them as the 1st, 2nd, and so on in the increasing order of their peak positions. Table S1 summarizes the parameter conditions of profile fitting. We carefully selected the analytical setup to achieve the good agreement between the experimental $I(Q)$ and calculated $I(Q)$, as well as the minimization of parameter cross-correlation. The asymmetry parameter $A$ are systematically fixed to 0.80 for GeSe, 0.88 for $(GeSe)_{50}(GeTe)_{50}$, and 0.96 for GeTe. The decay rates of the 1st peak were fixed to $R_l = 2.2$ and $R_h = 0.7$ for the 1st peak, while we also set $A = 1.5$, $R_l = 1.2$ and $R_h = 1.2$ for the 2nd peak. The parameters of the other profiles were $A = 1.0$ and $R_l = R_h = 1.2$ (i.e. the profile shapes are symmetric). We note that the widths of the 3rd peak were fixed to a certain value for each scan, which were determined from the preliminary analysis of $I(Q)$ collected at the high-pressure conditions. The widths become constant within error for both binary and pseudo-binary compositions at high pressure, and we used the averaged values for refinement. Figs. S1-S5 show the results of the profile fitting both in the pseudo-binary and in the binary systems, and Figure S6 represents the pressure dependence of the refined parameters. It is noteworthy that we performed profile fitting without assuming a structure model of crystals if sample was not crystallized (i.e. no Bragg peaks from sample in $I(Q)$).

**Subtraction of sample-cell contributions for pre-peak analysis**

Figure S7 shows the low-$Q$ range of the $I(Q)$ of amorphous $(GeSe)_{50}(GeTe)_{50}$. The several tiny peaks around 0.8 Å$^{-1}$ and around 1.4 Å$^{-1}$ originated from the Bragg reflections from the components of the high-pressure cells, which is demonstrated by the absence of peak shifts during compression. We estimated the contribution from a sample cell using $I(Q)$ at the highest-pressure point, where the intrinsic contribution from sample should be minimized. We performed linear interpolation (as shown in Figure S7), and the intensity profile of background was determined from the difference between $I(Q)$ and linear baseline. The extracted intensity profile of background was subsequently subtracted from the $I(Q)$, and we used the same profile for all the data in the same pressure scan. Figs. 2a-2c in the main text represent the $I(Q)$ after background subtraction.

**Table S1.** The summary of analysis setup of profile fitting. 10 profile functions are indexed in an increasing order for $2\theta$ positions. The columns *i*, *t*, *w*, *a* are the amplitude, position, width, and asymmetry parameters of the split-type Pearson-VII function. $R_l$, and $R_h$ represent the decay rates of peak intensity at the lower- side and the higher $2\theta$ sides of each peak.

| Peak Index | *i* | *t* | *w* | *a* | $R_l$ | $R_h$ |
|---|---|---|---|---|---|---|
| 1st | Refined | Refined | Refined | 0.80 (GeSe), 0.88 (Ternary), 0.96 (GeTe) | 2.20 | 0.70 |
| 2nd | | | Refined | 1.50 | 1.20 | 1.20 |
| 3rd | | | Fixed, the values determined from high-pressure results | 1.00 | | |
| 4th | | | Refined | | | |
| 5th | | | Refined | | | |
| 6th~10th | | | 5.00 | | | |

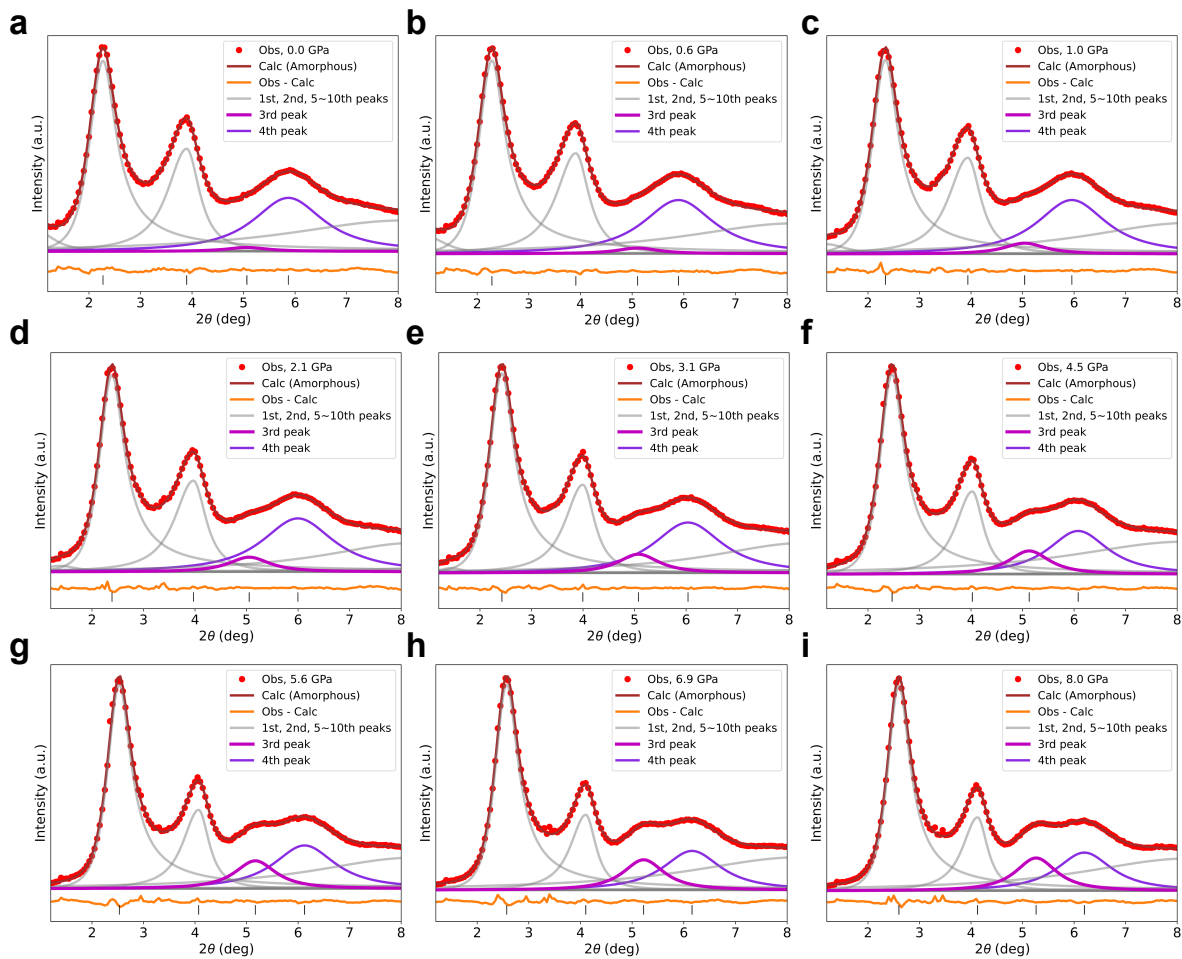

**Fig. S1 The results of the profile fitting of amorphous (GeSe)$_{50}$(GeTe)$_{50}$ at 300 K** (a)0.0 GPa, (b)0.6 GPa, (c)1.0 GPa, (d)2.1 GPa, (e)3.1 GPa, (f)4.5 GPa (g)5.6 GPa, (h)6.9 GPa, and (i)8.0 GPa. The red dots and brown line represent the observed and the calculated diffraction profiles, and the orange line show the difference between them. The magenta and purple lines highlight the contribution of the 3$^{rd}$ peak and the 4$^{th}$ peak, while the contributions from the other peaks were plotted as the gray lines.

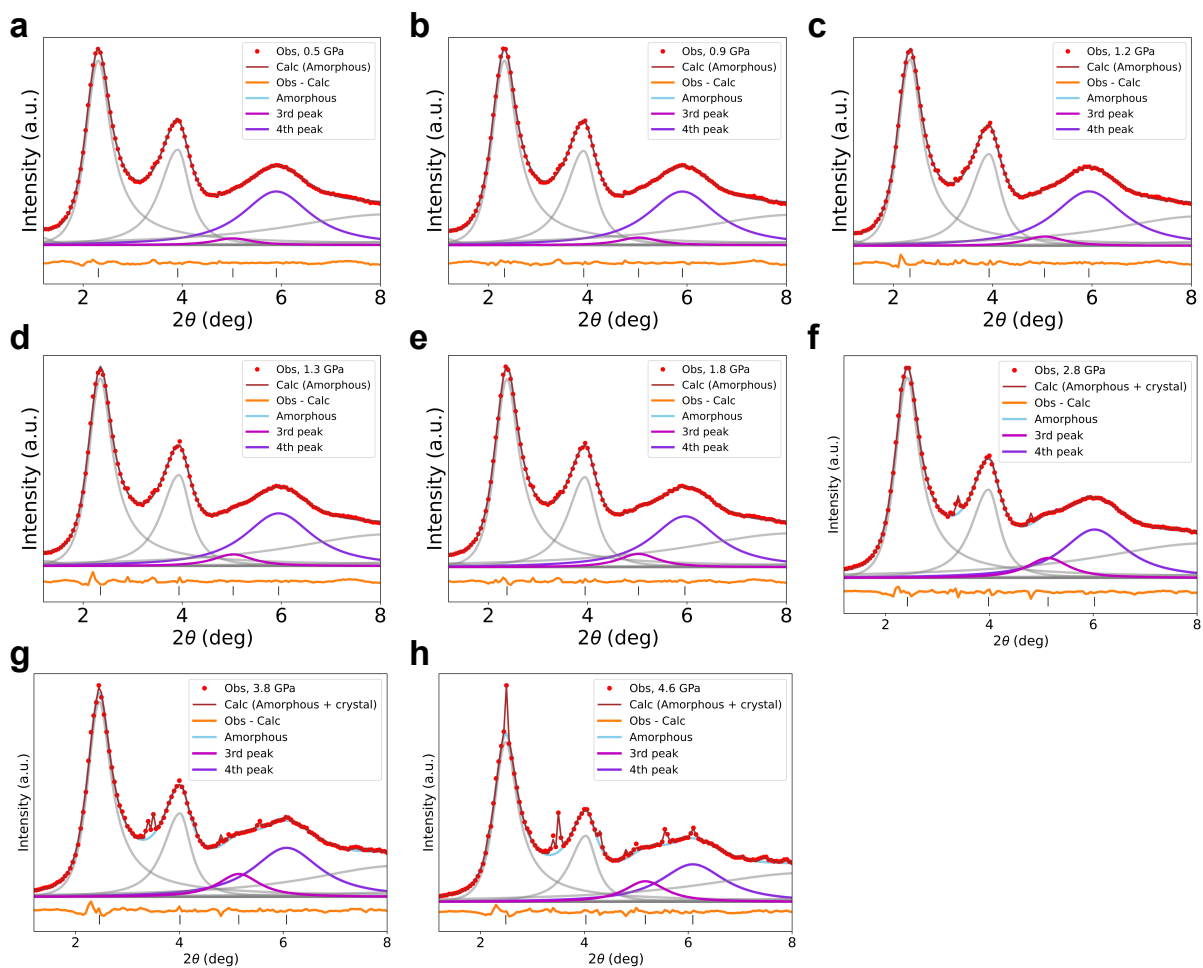

**Fig. S2 The results of the profile fitting of amorphous (GeSe)$_{50}$(GeTe)$_{50}$ at 373 K at** (a) 0.5 GPa, (b)0.9 GPa, (c)1.2 GPa, (d)1.3 GPa, (e)1.8 GPa, (f)2.8 GPa, (g)3.8 GPa, (h) 4.6 GPa. The red dots and brown line represent the observed and the calculated diffraction profiles, and the orange line show the difference between them. The magenta and purple lines highlight the contribution of the 3$^{rd}$ peak and the 4$^{th}$ peak, while the contributions from the other peaks were plotted as the gray lines. The sky-blue line represents the amorphous contributions as the sum of the 1$^{st}$ to the 10$^{th}$ peaks.

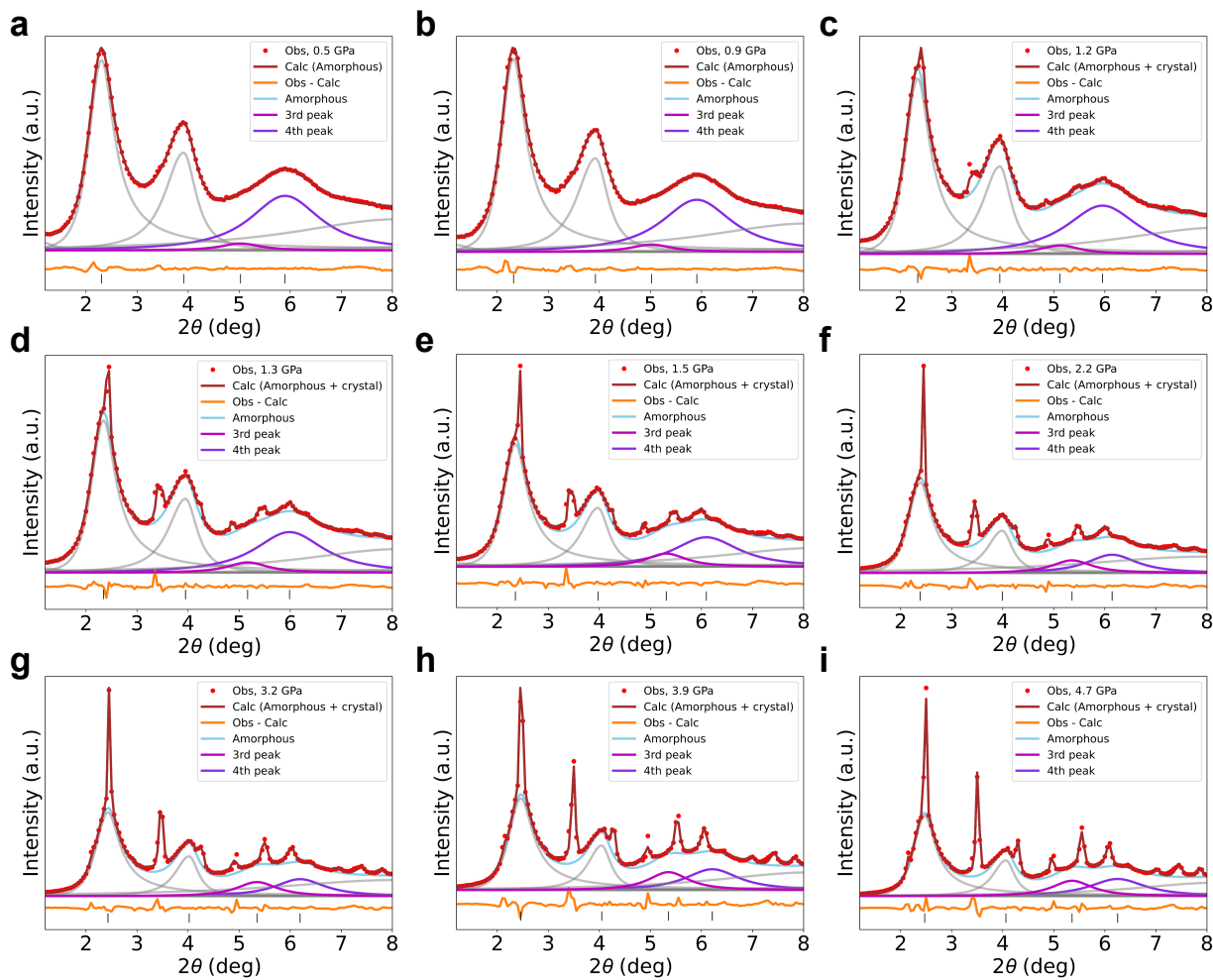

**Fig. S3 The results of the profile fitting of amorphous (GeSe)$_{50}$(GeTe)$_{50}$ at 423 K at** (a)0.5 GPa, (b)0.9 GPa, (c)1.2 GPa, (d)1.3 GPa, (e)1.5 GPa, (f)2.2 GPa, (g)3.2 GPa, (h)3.9 GPa, (i)4.7 GPa. The red dots and brown line represent the observed and the calculated diffraction profiles, and the orange line show the difference between them. The magenta and purple lines highlight the contribution of the 3$^{rd}$ peak and the 4$^{th}$ peak, while the contributions from the other peaks were plotted as the gray lines. The sky-blue line represents the amorphous contributions as the sum of the 1$^{st}$ to the 10$^{th}$ peaks.

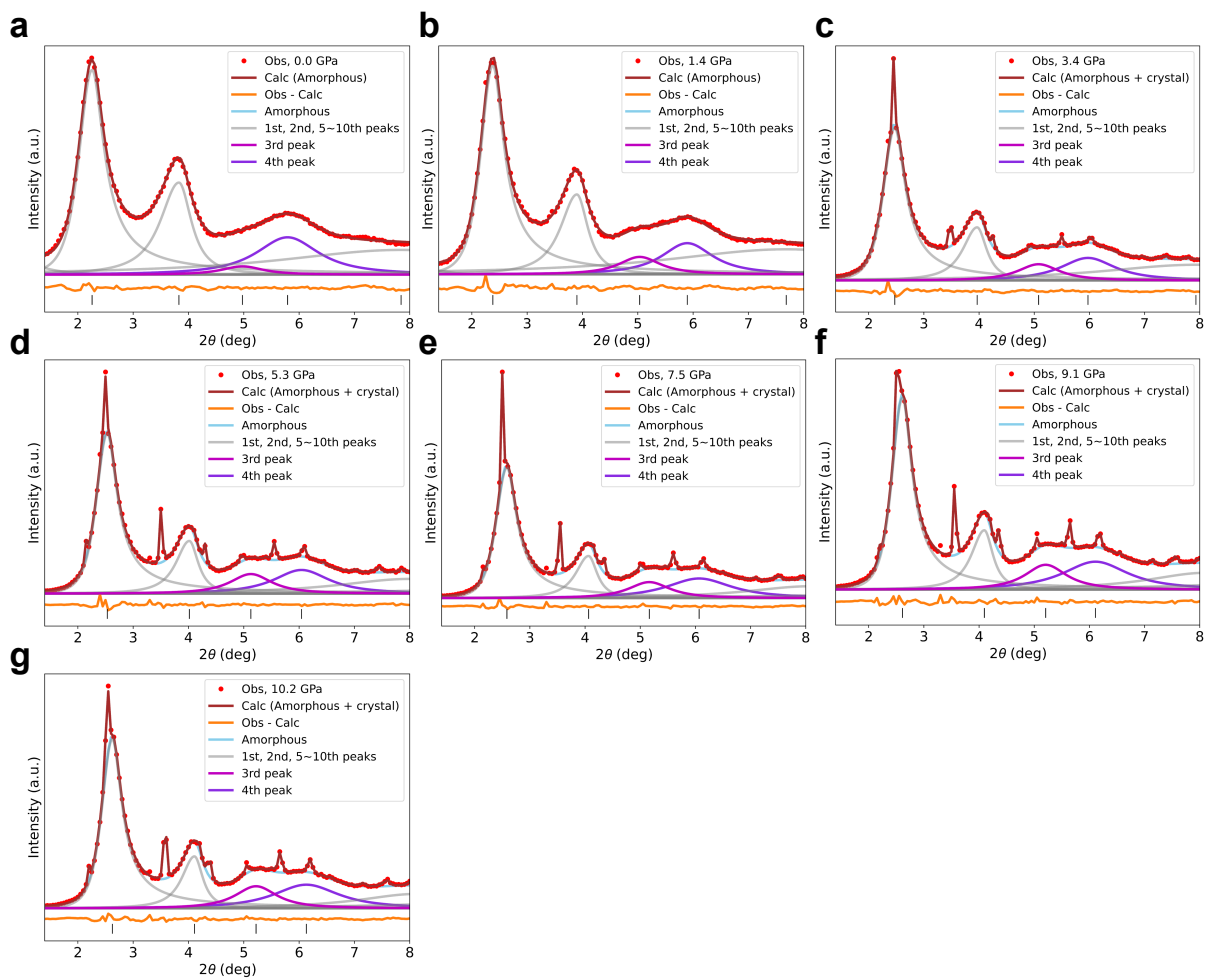

**Fig. S4 The results of the profile fitting of amorphous GeTe at ambient temperature** (a)0.0 GPa, (b)1.4 GPa, (c)3.4 GPa, (d)5.3 GPa, (e)7.5 GPa, (f)9.1 GPa (g)10.2 GPa. The red and brown data The red dots and brown line represent the observed and the calculated diffraction profiles, and the orange line show the difference between them. The magenta and purple lines highlight the contribution of the $3^{rd}$ peak and the $4^{th}$ peak, while the contributions from the other peaks were plotted as the gray lines. The sky-blue line represents the amorphous contributions as the sum of the $1^{st}$ to the $10^{th}$ peaks.

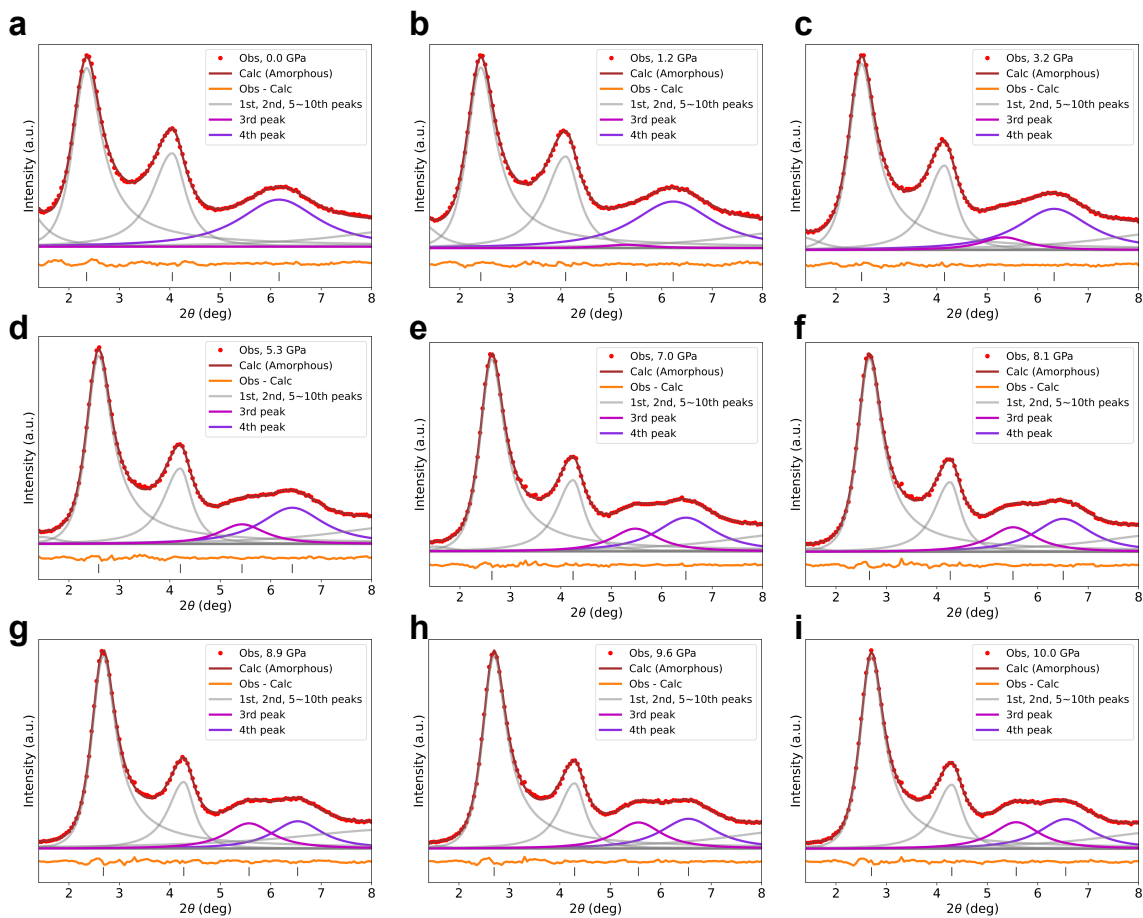

**Fig. S5 The results of the profile fitting of amorphous GeSe at ambient temperature** (a)0.0 GPa (b)1.2 GPa (c)3.2 GPa (d)5.3 GPa (e)7.0 GPa (f)8.1 GPa (g)8.9 GPa (h)9.6 GPa (i)10.0 GPa. The red dots and brown line represent the observed and the calculated diffraction profiles, and the orange line show the difference between them. The magenta and purple lines highlight the contribution of the 3[rd] peak and the 4[th] peak, while the contributions from the other peaks were plotted as the gray lines.

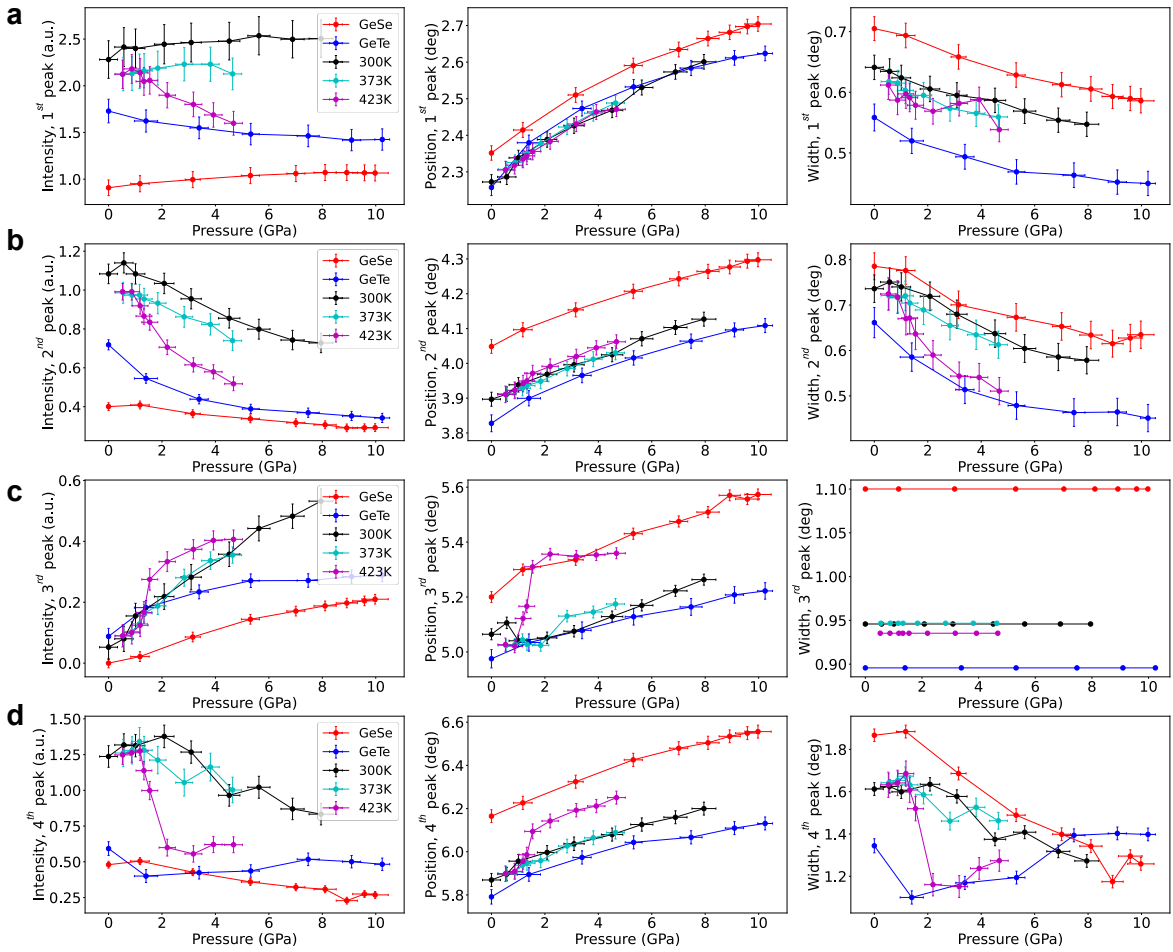

**Fig. S6 The refined parameters of the (a) 1ˢᵗ peak, (b) 2ⁿᵈ peak, (c) 3ʳᵈ peak, and (d) 4ᵗʰ peak of the profile fitting.** The pressure dependence of the amplitude, position, width from the left to the right. The parameter setup is summarized in Table S1..

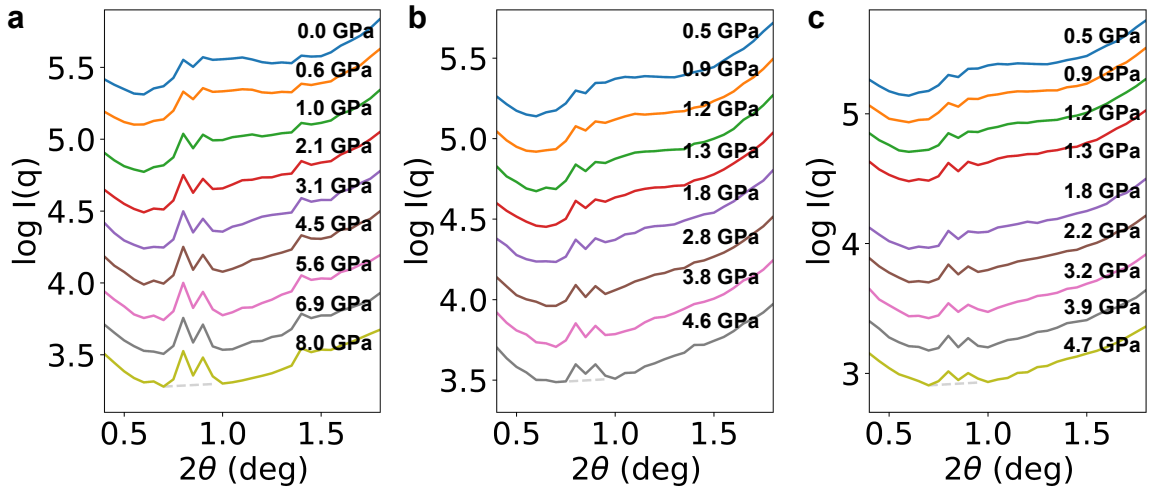

**Fig. S7. The raw diffraction profiles $I(Q)$ of amorphous (GeSe)$_{50}$(GeTe)$_{50}$** in the logarithmic scale at (**a**) 300 K, (**b**) 373 K, and (**c**) 423 K. The sharp peaks around 0.7° are the background Bragg peaks originated from the components of the high-pressure cells, as demonstrated by the absence of the peak shifts during compression. The gray dashed lines along with the $I(Q)$ at the highest-pressure point are the baselines to estimate the background intensity profiles.

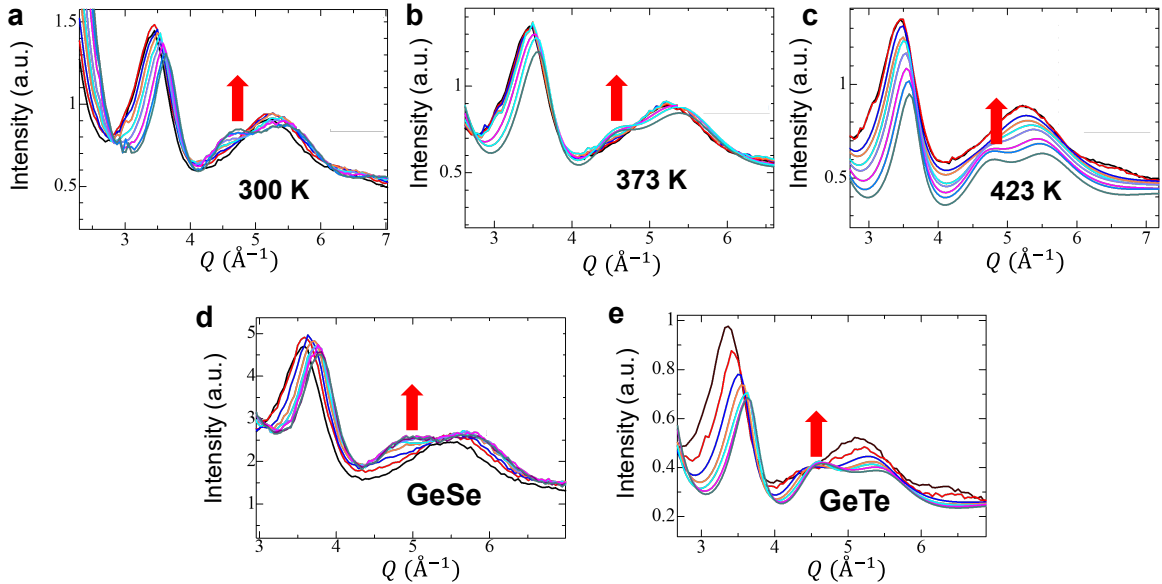

**Fig. S8** The pressure dependence of the diffraction profiles $I(Q)$ at (**a**) 300 K, (**b**) 373 K, (**c**) 423 K of the ternary (GeSe)$_{50}$(GeTe)$_{50}$, and (**d**) GeSe, and (**e**) GeTe. The medium $Q$-range from 2.5 Å$^{-1}$ to 7 Å$^{-1}$ are magnified. For the partially crystallized data, we plot the extracted amorphous contribution by profile fitting.

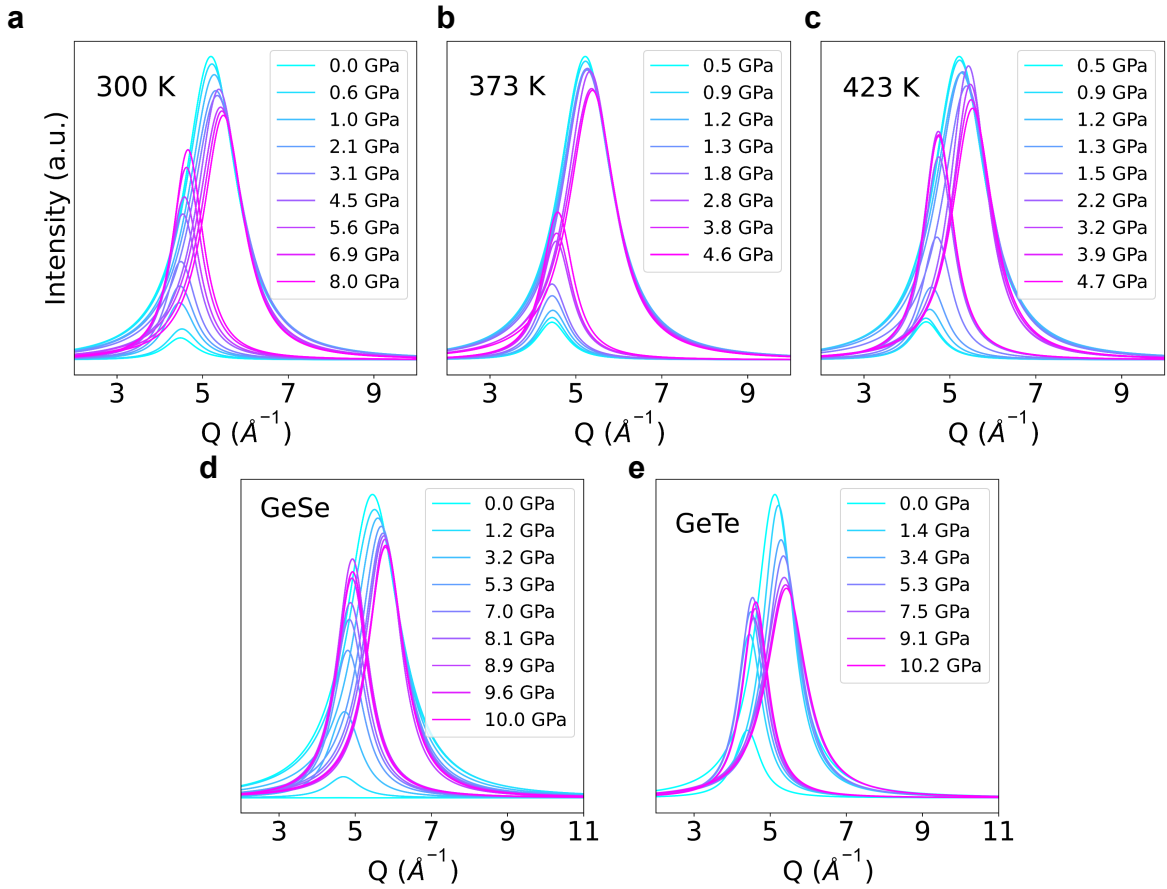

**Fig. S9 The pressure responses of the 3rd peak and the 4th peak during compression.** The amorphous $(GeSe)_{50}(GeTe)_{50}$ at (**a**) 300 K, (**b**) 373 K, (**c**) 423 K, and those of the binary alloys (**d**) GeSe and (**e**) GeTe. The pressure response of $(GeSe)_{50}(GeTe)_{50}$ at 423 K is distinguished from those of the other systems by the rapid rise of the 3rd peak and the pronounced sharpening of the 4th peak above 1.3 GPa. For comparison, intensity profiles at each pressure are normalized to the sum of the peak areas of the two peaks $i_3 + i_4$.

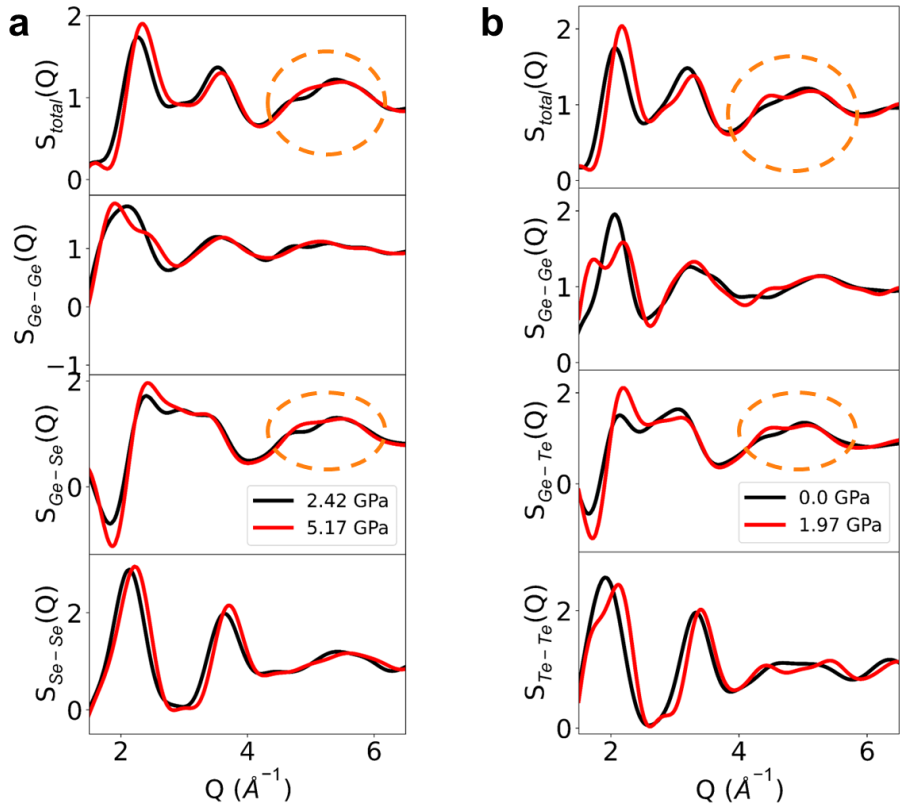

**Fig. S10** The total and partial structure factors $S(Q)$ of **(a)** amorphous GeSe and those of **(b)** amorphous GeTe, obtained from the ab-initio molecular dynamics simulation (AIMD) and the machine-learning molecular dynamics simulations with the neural-network potentials (NNMD), respectively. The partial structure factors show that the pressure-induced change in the Ge-Se and Ge-Te bond length distributions are responsible for the observed change of the $S_{\text{total}}(Q)$ around 4.5 Å$^{-1}$ to 4.7 Å$^{-1}$.

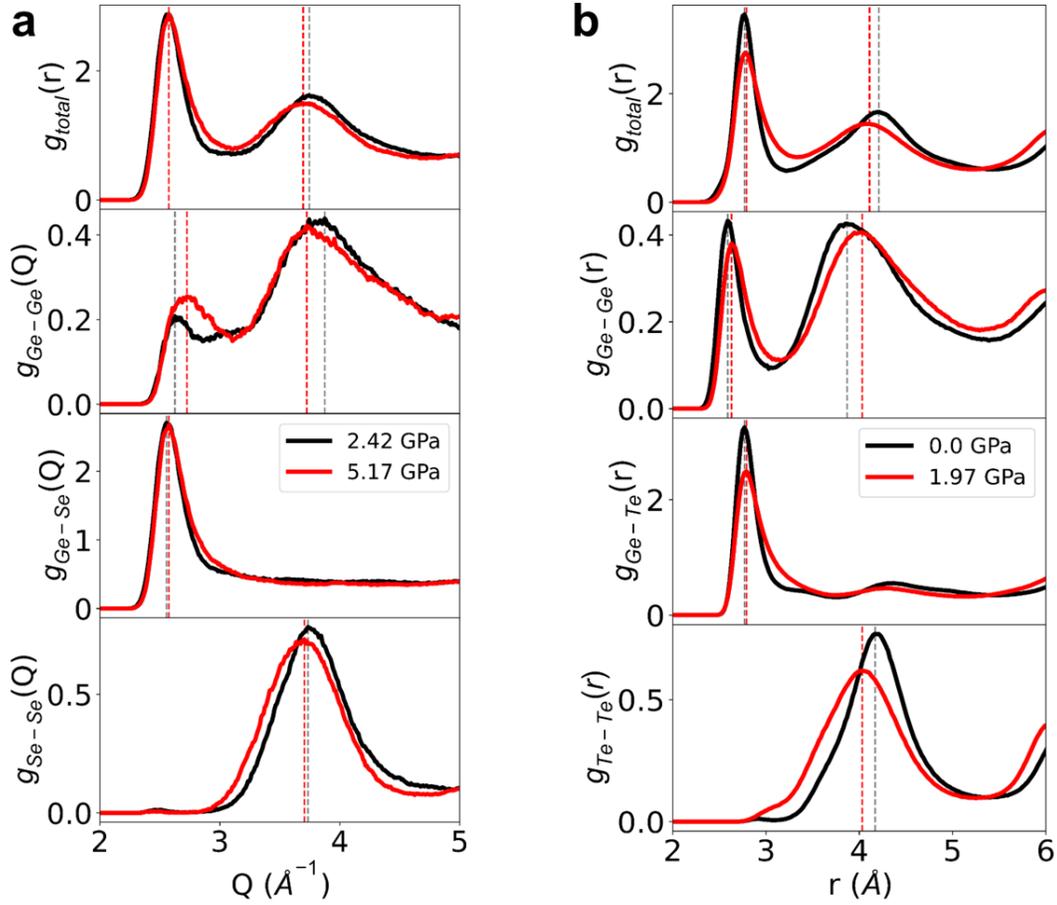

**Fig. S11** The total and partial pair distribution functions $g(r)$ of **(a)** amorphous GeSe and those of **(b)** amorphous GeTe, obtained from the ab-initio molecular dynamics simulation (AIMD) and the machine-learning molecular dynamics simulations with the neural-network potentials (NNMD), respectively. $g(r)$ show that the heteropolar Ge-Se (Te) bonds mainly contribute to the first shell, while Te(Se)-Te(Se) bonds occupy the second shell. The homopolar Ge-Ge bonds contribute to both the first and second shells. The vertical dashed lines represent the positions of the first two peaks by taking the local maximum of $g(r)$ from 2 Å to 3 Å and from 3 Å to 5 Å.

**Table S2.** Time lapses of pressure scan at 423 K. The exposure time was identical for all the data points. The standard error of time interval between two measurements was about 2 min.

| Pressure (GPa) | Exposure time (min) | Interval from last measurement (min) |
|---|---|---|
| 0.5 | | 22 |
| 0.9 | | 23 |
| 1.2 | | 22 |
| 1.3 | | 22 |
| 1.5 | 30 | 20 |
| 2.2 | | 25 |
| 3.2 | | 19 |
| 3.9 | | 25 |
| 4.7 | | 22 |
| Standard error | 0 | 2 |